\documentclass[sigconf, nonacm]{acmart}

\usepackage{graphicx}

\usepackage{algorithm}[1]
\usepackage{algpseudocode}

\usepackage{amsmath,amsfonts}
\usepackage{subfig}
\usepackage{multirow}
\usepackage{array}

\usepackage[english, status=draft, margin=false, inline=true]{fixme}
\fxusetheme{color}
\FXRegisterAuthor{jf}{ajf}{\color{blue}JF}
\FXRegisterAuthor{cg}{acg}{\color{orange}CG}

\newcommand{\rmspace}{\vspace{-0.5ex}}

\begin{document}

\title{Interactive Ontology Matching with Cost-Efficient Learning}

\author{Bin Cheng}
\affiliation{%
  \institution{NEC Laboratories Europe}
}
\affiliation{%
  \institution{Springer Nature}
}
\email{bin.cheng@springernature.com}

\author{Jonathan F{\"u}rst}
\affiliation{%
  \institution{Zurich University of Applied Sciences}
}
\affiliation{%
  \institution{NEC Laboratories Europe}
}
\email{jonathan.fuerst@zhaw.ch}

\author{Tobias Jacobs}
\affiliation{%
  \institution{NEC Laboratories Europe}
}
\email{tobias.jacobs@neclab.eu}

\author{Celia Garrido-Hidalgo}
\affiliation{%
  \institution{Universidad de Castilla-La Mancha}
}
\email{celia.garrido@uclm.es}

\begin{abstract}

The creation of high-quality ontologies is crucial for data integration and knowledge-based reasoning, specifically in the context of the rising data economy. However, automatic ontology matchers are often bound to the heuristics they are based on, leaving many matches unidentified.
Interactive ontology matching systems involving human experts have been introduced, but they do not solve the fundamental issue of flexibly finding additional matches outside the scope of the implemented heuristics, even though this is highly demanded in industrial settings.
Active machine learning methods appear to be a promising path towards a flexible interactive ontology matcher.
However, off-the-shelf active learning mechanisms suffer from low query efficiency due to extreme class imbalance, resulting in a \emph{last-mile problem} where high human effort is required to identify the remaining matches.

To address the last-mile problem, this work introduces \emph{DualLoop}, an active learning method tailored to ontology matching. DualLoop offers three main contributions: (1) an ensemble of tunable heuristic matchers, (2) a short-term learner with a novel query strategy adapted to highly imbalanced data, and (3) long-term learners to explore potential matches by creating and tuning new heuristics. We evaluated DualLoop on three datasets of varying sizes and domains. Compared to existing active learning methods, we consistently achieved better F1 scores and recall, reducing the expected query cost spent on finding 90\% of all matches by over 50\%. Compared to traditional interactive ontology matchers, we are able to find additional, last-mile matches. Finally, we detail the successful deployment of our approach within an actual product and report its operational performance results within the Architecture, Engineering, and Construction (AEC) industry sector, showcasing its practical value and efficiency.
\end{abstract}

\maketitle

\section{Introduction}

Ontologies are an established way of 
describing classes, properties, and relationships 
in the form of triples, representing knowledge that can then be reused and shared across 
systems and domains.
Ontologies play an important role in data integration~\cite{gao2015data} and 
automated 
reasoning~\cite{hohenecker2020ontology}.
Lately, ontologies serve as the backbone of knowledge graphs~\cite{dong2023generations}, as interoperability mechanism in data spaces and marketplaces~\cite{solmaz2022enabling}, and have given rise to a plethora of applications of graph-based machine learning~\cite{xia2021graph}. 
Concretely, 
enterprises 
leverage 
ontologies 
to align the 
schemata 
of their heterogeneous data sources into a common knowledge graph for 
more effective
information sharing
between departments, as well as for assisting the development of downstream data analytics and reasoning tasks like e.g. question-answering~\cite{kosten2023spider4sparql}. In data spaces and marketplaces, the use of ontologies as interoperability mechanism is seamlessly extended across organizational borders.
The centrality of ontologies in all these scenarios has brought new attention~\cite{bento2020ontology, maresca2021ontoaugment, furst2023versamatch} to the time-honored topic of ontology matching---GLUE~\cite{doan2003learning}, one of the seminal works on automatic matching has been published by the database community more than 20 years ago!

\begin{figure}[ht]
    \centering
    \includegraphics[width=1.0\linewidth]{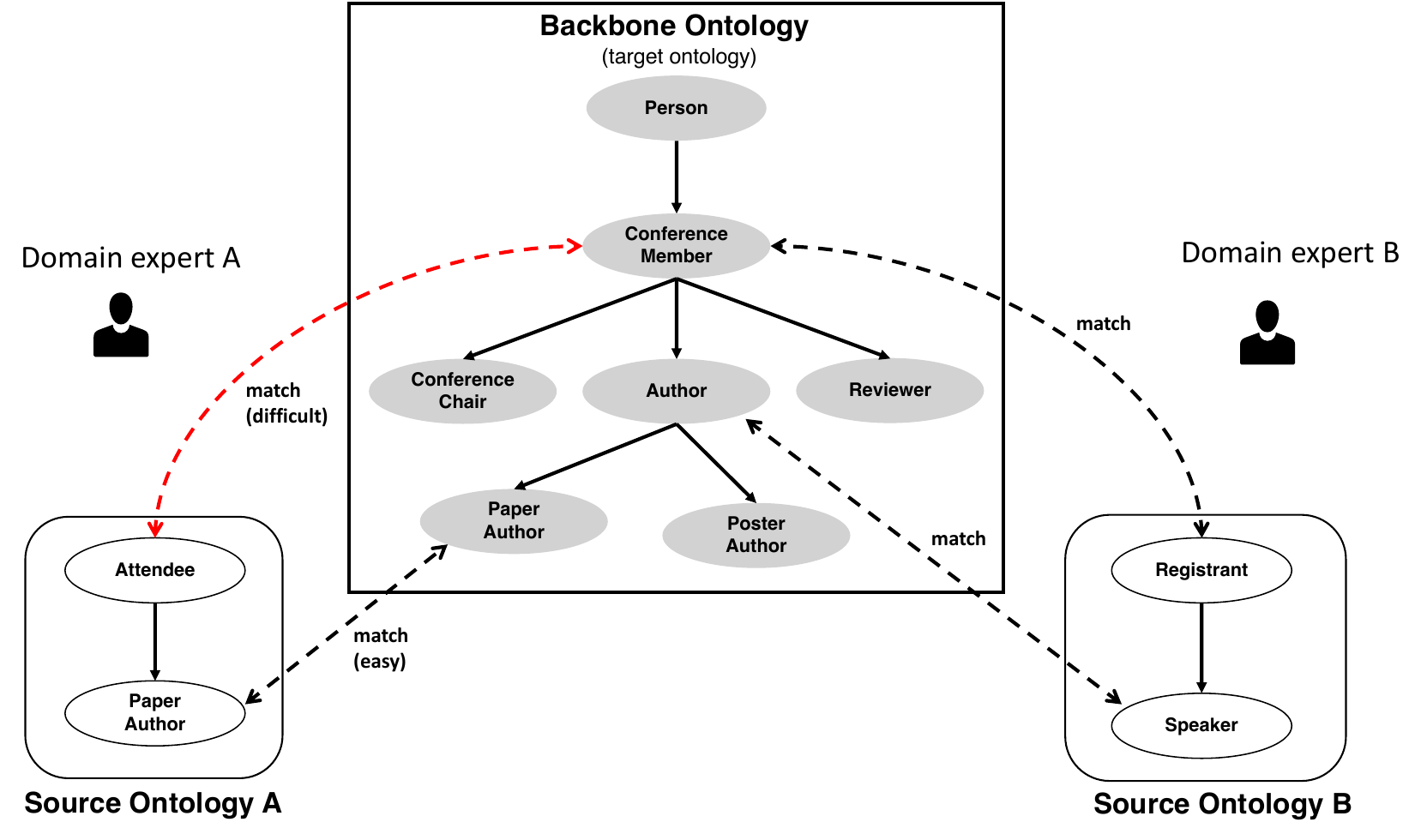}
    \caption{Aligning different source ontologies with a backbone ontology to facilitate interoperability. Domain experts are available to support an automated system.}
    \label{fig:backbone-ontology}
    \rmspace
\end{figure}

\emph{Ontology matching}~\cite{shvaiko2011ontology},
also called~\emph{ontology alignment}, aims to identify inter-ontology relationships (e.g. equivalence, subsumption) between pairs of elements (classes, relations, properties) from two given ontologies. This task has the potential to be automated to a large extent, helping 
domain experts to speed up 
the merging process as illustrated in Figure~\ref{fig:backbone-ontology}. Here, a backbone ontology is used to achieve interoperability between two different source domains. Domain experts are involved in the alignment process. This is the concept of \emph{NEC's TrioNet toolkit} (see Section~\ref{subs:trionet}). In this usage scenario, the \emph{key success criteria for ontology matchers is a high recall (finding all matches) combined with a high precision (not miss-classifying matches)}. The \emph{human annotation efforts are often secondary} and companies are willing to go the extra mile to achieve a wider integration.

\begin{figure}[ht]
    \centering
    \includegraphics[width=1.0\linewidth]{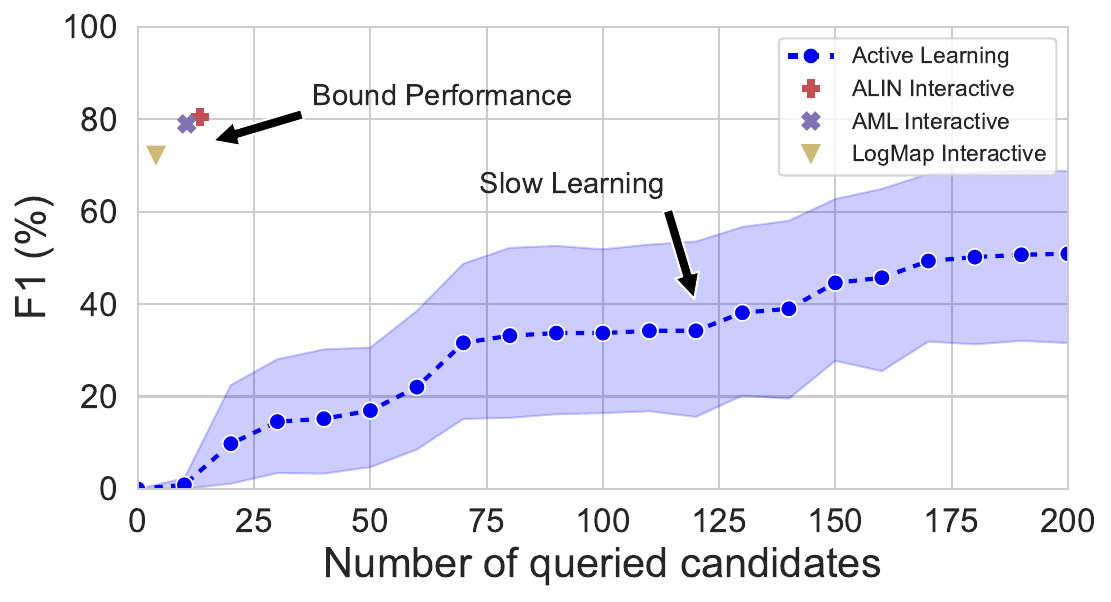}
    \caption{Problems with existing interactive matchers and active learning on the Conference benchmark dataset~\cite{OAEI-Conference}: interactive ontology matchers allow for human-in-the-loop user annotations, but their performance is limited, and users have no option to provide further annotations to improve performance beyond a certain point; active learning exhibits an extremely low sample efficiency, impractical for real applications. }
    \label{fig:f1_intro}
    \rmspace
\end{figure}

Existing 
fully automatic 
ontology matchers, such as AML~\cite{faria2013agreementmakerlight}, LogMap~\cite{jimenez2011logmap}, Yam++~\cite{ngo2012yam}, and VersaMatch~\cite{furst2023versamatch} 
are able to 
predict a 
subset of matches
with high precision, 
but the results are restricted to the reach of their implemented matching heuristics, leaving many matches unidentified.
To overcome the limitations of 
fully automated matching, 
others have added an \emph{interactive, human-in-the-loop} step to 
help traditional
fully automatic matchers~\cite{faria2013agreementmakerlight,jimenez2011logmap,ALIN2020}.
While this improves upon the automatic results~\cite{OAEI-Interactive}, it does not 
overcome the fundamental issue that exploration of matches is restricted to the hard-coded 
matching heuristics. Thus, user interaction is limited to refining the obtained results, e.g., by verifying class pairs that are close to the decision boundary. This bound performance prohibits a discovery of last mile matches (see left top corner of Figure~\ref{fig:f1_intro}).

A general and well-studied interactive approach, which has the advantage of full flexibility in the number of queries submitted to the human expert and in its learning capabilities, is active learning~\cite{shi2009actively}. Active learning based systems have shown great success for related classification problems (e.g., for schema alignment~\cite{ten2018active}
and entity matching~\cite{meduri2020comprehensive})
However, we experimentally show that, for ontology matching tasks, existing active learning algorithms
(1) struggle to bootstrap their learning capability due to their uninformed initial selection strategy
and (2) exhibit a bad precision-recall trade-off (see blue curve with circle markers in Figure~\ref{fig:f1_intro}).
This problem is particularly severe 
when dealing with the extreme class imbalance 
of the ontology matching task, where typically far less than~1\% of all class pairs correspond to true matches. 
As a result, existing active learning approaches suffer from a low \emph{query efficiency}, presenting many non-matches to the human expert to overcome the precision-recall trade-off.

In this work, we address the identified niche between existing precise and efficient, yet constrained (interactive) ontology matchers, and the more flexible but less efficient active learning paradigm.
\emph{DualLoop} employs recent weak supervision methods from data programming~\cite{ratner2020snorkel} for approximate but automatic data labeling. To tackle the unfavorable precision-recall trade-off in active learning, we introduce a novel dual-query strategy, which consists of a short-term learner emphasizing exploitation (i.e., enhancing precision) and long-term learners facilitating exploration (i.e., boosting recall).
The result is a unique ontology matcher that offers unbound capabilities to find additional, last-mile matches with a high query efficiency to minimize human effort---a requirement urgently needed in practical applications of ontology matching.
Summarized, the high query efficiency of DualLoop is the result of three main technical contributions:
\let\labelitemi\labelitemii
\begin{itemize}
\item We employ an ensemble of \emph{weak labeling functions}, each assigning noisy class labels (match or no-match) to a subset of the class pairs. The ensemble bootstraps the active learning process, and is responsible for the solid initial performance. 
Later, when human annotations have been collected, those true annotations are utilized to filter and weight functions within the ensemble.

\item Our \emph{short-term learner} systematically selects high-confidence matching candidates identified by the labeling function ensemble. Prioritizing exploitation in that way turns out successful to overcome the challenge of the extreme class imbalance, inherent to the ontology matching task. 

\item To explore the space of potential matches beyond the initial labeling functions, \emph{long-term learners} automatically create and configure \emph{tunable labeling functions} that are based on a variety of distance metrics. These functions are tuned for potential generalization from the set of matches that have been verified during the active learning process.
\end{itemize}

We evaluate the 
query
efficiency of DualLoop 
by applying it to three datasets covering different domains. 
Our experiments demonstrate that
DualLoop  consistently attains higher F1 scores and recall compared to other, existing flexible active learning methods,
while reducing the expected query cost needed to discover 90\% of all matches by over 50\%. 
In comparison to state-of-the-art interactive ontology matchers, DualLoop effectively uncovers additional matches, aligning with the level of effort users are willing to invest to walk the last mile.

Last, we present our implementation of Dualloop in a data interoperability system called TrioNet that has been developed for a data sharing platform at NEC and been transferred from research prototype to platform product. To show its practical applicability, we use TrioNet to match ontologies found in the Architecture, Engineering, and
Construction (AEC) industry and compare the required user interactions and discovered matches to state of the art matchers.   
\section{Preliminaries}
\label{sec:problem}

\subsection{Ontology Matching}

An ontology is a formal conceptualization of a domain with
classes, instances, properties, as well as relations between them. 
The latter includes fundamental relations like \textit{superclass}, \textit{subclass}, and \textit{is-instance-of}, defining a hierarchical structure with inheritance of properties. 

Given two ontologies $S$ and $T$ with associated sets of classes $E_S$ and $E_T$, a \emph{class match} is an inter-ontology relation $(e_S,r,e_T)$, where $e_S \in E_S$, $e_T \in E_T$, and $r$ is the specific type of relation between the two classes. While $r$ can possibly represent any such relation type, including e.g. subsumption or disjointness, the focus of our work is on the most commonly addressed problem of identifying class equivalence relations. Hence, we consider ontology matching as a binary classification task characterized by 
$$\{(e_S,e_T) \mid e_S \in E_S \textrm{ equivalent to } e_T \in E_T  \}$$ 
as the set of elements with class label $1$.

\subsection{Interactive Ontology Matching}

Misclassification costs in ontology matching are highly asymmetrical. The reason is that, in the downstream data integration process, false negatives mostly lead to some degree of data redundancy, whereas false positives can cause data loss during automated removal of redundancies. Likewise, automated reasoning engines might miss some conclusions due to false negatives, while false positive matches can cause wrong conclusions. For these reasons, it is often required in practice that machine-identified matches are \emph{verified} by a human domain expert. The limitations of fully automated matching approaches discussed in the preceding section provide additional motivation for \emph{interactive ontology matching}. The idea is to make the involvement of human experts an integral part of the matching algorithm and learn from the feedback, instead of presenting the set of identified matches only at the end of a fully automated process. 

In our experiments we employ two methods to evaluate the performance of active learning algorithms for ontology matching. The first method is to put binary classification metrics (F1-score) in relation to the number of queries to the \emph{oracle} (human expert). This is a standard metric for active learning. The other method is based on the assumption that all matches must be verified by a domain expert. Here the query cost is calculated by summing up the number of annotations during the active learning phase and the number of matches predicted by the final model. This cost is put in relation to the recall, that is, the percentage of true matches identified and verified.
We also report the \emph{response time}, which is defined as the wall clock time that passes between the reception of an answer from the oracle and the submission of the subsequent query.

\subsection{Weak Supervision}

The weak supervision paradigm has been established to make supervised machine learning methods applicable in face of absence of labeled data. We adapt the idea of data programming~\cite{ratner2016data} and rely on a family $\Lambda$ of simple and noisy \emph{labeling functions}. In the context of the ontology matching problem,  each $\lambda:E_S \times E_T \to \{0,1\}, \lambda \in \Lambda$, is a partial function that labels a subset of all class pairs as being either no match or a match. We remark that in our system the family of labeling functions is not fixed. but it is dynamically enhanced to explore further categories of matches; details will be described in the subsequent section.

As in this work we deal with a combination of true class labels (assigned by oracle interaction) and weak labels (assigned by labeling functions), we refer to the former as \emph{annotations} and to the latter as \emph{labels}. 
\section{Methodology}

Our methodology, named \emph{DualLoop}, combines ideas from active learning with weak supervision.

Due to the large number of potential matches, which is quadratic in the size of the input, it is common in ontology matching to apply an initial filtering function - often called \emph{blocking} - to reduce the number of matching candidates~\cite{thirumuruganathan2021deep}. Our system makes use of a key-based blocking method and the detail is introduced in Section~\ref{sec:blocking}

The role of the labeling functions is to form a \emph{voting committee} that provides estimations of the probability of any given class pair to be a match. This information is used, during the active learning phase, to select potential matches for the oracle to annotate. Conversely, the already annotated data is utilized to curate the committee and prioritize the votes. The mechanism just described is called the \emph{fast loop}, because it is executed between subsequent interactions with the oracle and must therefore have a short runtime.  A pseudo-code description is provided in Algorithm~\ref{alg:fastloop}.

The \emph{slow loop} is executed in parallel to the fast loop; its purpose is  
to augment 
the family of labeling functions 
with automatically created and tuned labeling functions
for exploring the space of possible matches 
beyond the reach of the initial set. 
Both the annotated data and the voting committee are employed for parameter tuning of the additional labeling functions; 
see Algorithm~\ref{alg:slowloop} for pseudo-code.

In the following subsections, we provide technical details of the query strategy and the ensemble of labeling functions, which together form the fast loop. After that, we describe how in the slow loop the new labeling functions are tuned. 

\subsection{Query Strategy}

Let $X \subset E_S \times E_T$ be the set of initial matching candidates after application of the blocking method. Further, let $A \subset X$, initially empty, be the set of already annotated class pairs, and let $U := X \setminus A$. The query strategy selects one pair $(e_S,e_T) \in U$ to be presented to the oracle for annotation. 
The challenge is to use the annotation budget carefully to balance the opportunity of 
gaining new insights (exploration) and maximizing the number of verified matches (exploitation). 
Several query methods have been proposed in literature, 
such as sample uncertainty and query-by-committee~\cite{schein2007active}.
However, they turn to be inefficient
when dealing with the extreme class imbalance in ontology matching~\cite{mussmann2020importance}. 
DualLoop introduces a more efficient query strategy 
that prioritizes exploitation as long as the ensemble of labeling functions identifies high-confidence matches, 
while exploration happens through new labeling functions that are created by the slow loop. 

Let $\Lambda$ be the current set of labeling functions, and, for $\lambda \in \Lambda$, let $\hat{y}^\lambda_{u} := \lambda(u) \in \{0,1,-1\}$ the label predicted by $\lambda$ for a given class pair $u \in U$, where result $-1$ symbolizes that $u$ is outside the domain of $\lambda$. Further, let function $cmt:U \to \{0,1\} \times [0,1]:u \mapsto (\hat{y}^{cmt}_u,p^{cmt}_u)$ represent the current committee of labeling functions, which assigns to each unlabeled class pair $u\in U$ a predicted result (match or no match) $\hat{y}^{cmt}_u$ as well as a score $p^{cmt}_u$. The latter is a number between $0$ and $1$, representing the certainty of the committee that $u$ is a match. 

Our query strategy,
executed in line~8 of Algorithm~\ref{alg:fastloop},
breaks the sample query process into two levels of selection. 
First, all samples $u \in U$ are grouped based on 
$|\{\lambda(u)=1 \mid \lambda \in \Lambda\}|$, 
the number of 
labeling functions predicting a match, 
and the group with the highest number of predicted matches is selected. 
Second, within the selected group, 
the sample $u$ with the maximum certainty $p^{cmt}_u$ is picked for annotation. 
For computational efficiency reasons, we submit the queries to the oracle in batches of size $B$, repeating the sample selection process $B$ times before executing the batch.

\subsection{Ensemble of Labeling Functions} 

The set $\Lambda$ of labeling functions forms the basis for the voting committee, which provides, for any unlabeled class pair $u \in U$, a result $\hat{y}^{cmt}_u \in \{0,1\}$ as well as a score $p^{cmt}_u \in [0,1]$. 

Our method,
executed in line~6 of Algorithm~\ref{alg:fastloop},
is inspired by the data programming approach of Snorkel~\cite{ratner2020snorkel}, a framework designed to train classifiers from weak labeling functions, each assumed to have an unknown accuracy strictly higher than pure chance. 
We adapt the ensemble mechanism of DualLoop to the somewhat different situation where a small set $A$ of annotated data is available through active learning. Since $A$ enables us to estimate the performance of each labeling function, 
we can lift the a-priori assumption on the function quality,
an assumption hard to guarantee in practice.

First, we 
select a subset $\Lambda^\prime \subset \Lambda$ based on the annotated samples. Let $\textrm{F1}_A(\lambda)$ be the F1 score achieved by labeling function $\lambda$ on the samples in $A$, and let $\lambda_1,\ldots,\lambda_m$ be a sorting of the functions such that $\textrm{F1}_A(\lambda_1) \geq \ldots \geq \textrm{F1}_A(\lambda_m)$. Define $\Lambda_{j} := \{\lambda_1,\ldots,\lambda_j\}$, and let $\textrm{F1}_A(\Lambda_j)$ be the F1-score achieved by the voting committee composed of $\Lambda_j$ (the voting mechanism will be described below). Let $k$ be the largest integer such that $k \geq 3$ and $\textrm{F1}_A(\Lambda_{j-1}) < \textrm{F1}_A(\Lambda_{j})$ for each $3<j\leq k$. Then the final committee consists of $\Lambda^\prime := \Lambda_k$. 

To build a voting committee from a given function set $\Lambda^\prime$, each member is assigned a weight. For $\lambda \in \Lambda^\prime$ we compute $\textrm{prec}_A(\lambda)$, the precision w.r.t. set $A$. Then the weight $w_\lambda$ of $\lambda$ is computed as $w_0\cdot\textrm{prec}_A(\lambda)$ where $w_0:=0.7$ is a constant. The labeling functions from $\Lambda^\prime$ and their respective weights are then used as input to the ensemble fitting function implemented by Snorkel (see Section 2.2 in \cite{ratner2020snorkel}).

\begin{algorithm}[!t]

\caption{Fast loop}
\label{alg:fastloop}

\begin{algorithmic}[1]
\State $\Lambda \leftarrow \ \textrm{initial labeling functions}$
\State $cmt \leftarrow \Lambda \ \textrm{with uniform weights}$
\State $U \leftarrow \ \textrm{all class pairs after blocking}$
\State $A \leftarrow \emptyset$
\Repeat
    \If{$|A| \geq a_\textrm{min}$} update $cmt$ from $\Lambda$ according to $A$
    \EndIf
    \State use $cmt$ to select query set $Q \subset U$ of size $B$
    \State query oracle with $Q$
    \State $A \leftarrow A \cup Q, U \leftarrow U \setminus Q$
\Until{query budget exhausted}
\end{algorithmic}
\end{algorithm}
\rmspace

\subsection{Ensemble Augmentation}

The initial labeling functions provided by domain experts usually have a simple logic and are easy to program. 
Their performance often is conservative, i.e., they cover few cases with high certainty and use safe distance thresholds. 
The fast loop prioritizes exploitation of such highly likely matches. To further explore the space of possible matches, we introduce the slow loop to augment the ensemble with newly generated labeling functions. 
This helps to break the deadlock situation where the existing committee runs out of matching candidates, while randomly searching for further matches is not an option because of the high class imbalance.

We introduce the concept of \emph{tunable labeling functions} to create and/or update such functions on the fly. Let $D$ be a set of distance metrics $d:X \to \mathcal{R}^+$, assigning a non-negative distance $d(x)$ to each class pair $x \in X$. Each tunable labeling function $\lambda_{d,h}$ is characterized by a metric $d \in D$ and threshold $h>0$:
$$
    \lambda_{d,h}(x) = 
    \begin{cases}
        0 \ \ \textrm{if}\ d(x) > h \ , \\
        1 \ \ \textrm{if}\ d(x) \leq h \ .
    \end{cases}
$$
$D$~can include simple metrics like edit distance between the class attributes. Further, $D$ can contain advanced metrics like the pre-trained sentence transformer model Sentence-BERT~\cite{reimers2019sentence}, measuring the semantic similarity of string-based class attributes. Lastly, the matching probabilities predicted by machine learning models (such as random forest, logistic regression, MLP), trained from the latest set $A$ of labeled data and/or the voting committee predictions, can be included into $D$. 

For any metric $d \in D$, DualLoop maintains a maximum threshold $h_d^\textrm{max}$, which is initially zero. When tuning the threshold for $d$, the first step is to check whether the current committee predicts any matches for the current set $U$ of  class pairs not yet annotated. If not, then $h_d^\textrm{max}$ is increased by a constant $\delta$ for more generous exploration. After that, the threshold $h_d$ is selected, in line~8 of Algorithm~\ref{alg:slowloop}, by
$$
    h_d := \textrm{arg max}_{h \leq h_d^\textrm{max}} \bigl(\textrm{TP}^A_{d,h} + \textrm{prec}^{cmt}_{d,h} + \frac{B}{\max\{\textrm{FP}^{cmt}_{d,h},B\}}\bigr) \ ,
$$
where $B$ is the batch size and $\textrm{TP}^A_{h,d}$ is the number of true positives in $A$ identified by the labeling function $\lambda_{d,h}$. Further, $\textrm{prec}^{cmt}_{d,h}$ and $\textrm{FP}^{cmt}_{d,h}$ is the precision and the number of false positives, respectively,  of $\lambda_{d,h}$ measured against the voting results $cmt(u)$ for each $u \in U$. The first summand is a positive integer and represents the main criterion for tuning $h$ towards exploration. The other two summands take fractional values between $0$ and $1$ and serve as tie breakers. The term $\textrm{prec}^{cmt}_{d,h}$ rewards agreement with the committee about predicted matches, and the final term is a soft limit on the number of new matches proposed in disagreement with the committee. 
\begin{algorithm}
\caption{Slow loop}
\label{alg:slowloop}

\begin{algorithmic}[1]
\State $D \leftarrow \ \textrm{given distance metrics}$
\ForAll{$d \in D$ in parallel}
\State $h^\textrm{max}_d \leftarrow 0$ 
\State $\Lambda \leftarrow \Lambda \cup \{\lambda_{d,0}\}$
\Repeat
    \If{\text{$cmt$ predicts no match}} $h^\textrm{max}_d \leftarrow h^\textrm{max}_d+\delta$
    \EndIf
    \State select threshold $h_\textrm{new} \leq h^\textrm{max}_d$ based on $A$ and $cmt$
    \State replace $\lambda_{d,\_}$ in $\Lambda$ with $\lambda_{d,h_\textrm{new}}$
\Until{query budget exhausted by fast loop}
\EndFor
\end{algorithmic}
\end{algorithm}
\rmspace  
\subsection{DualLoop System}

\begin{figure*}
    \centering
    \includegraphics[width=1.0\linewidth]{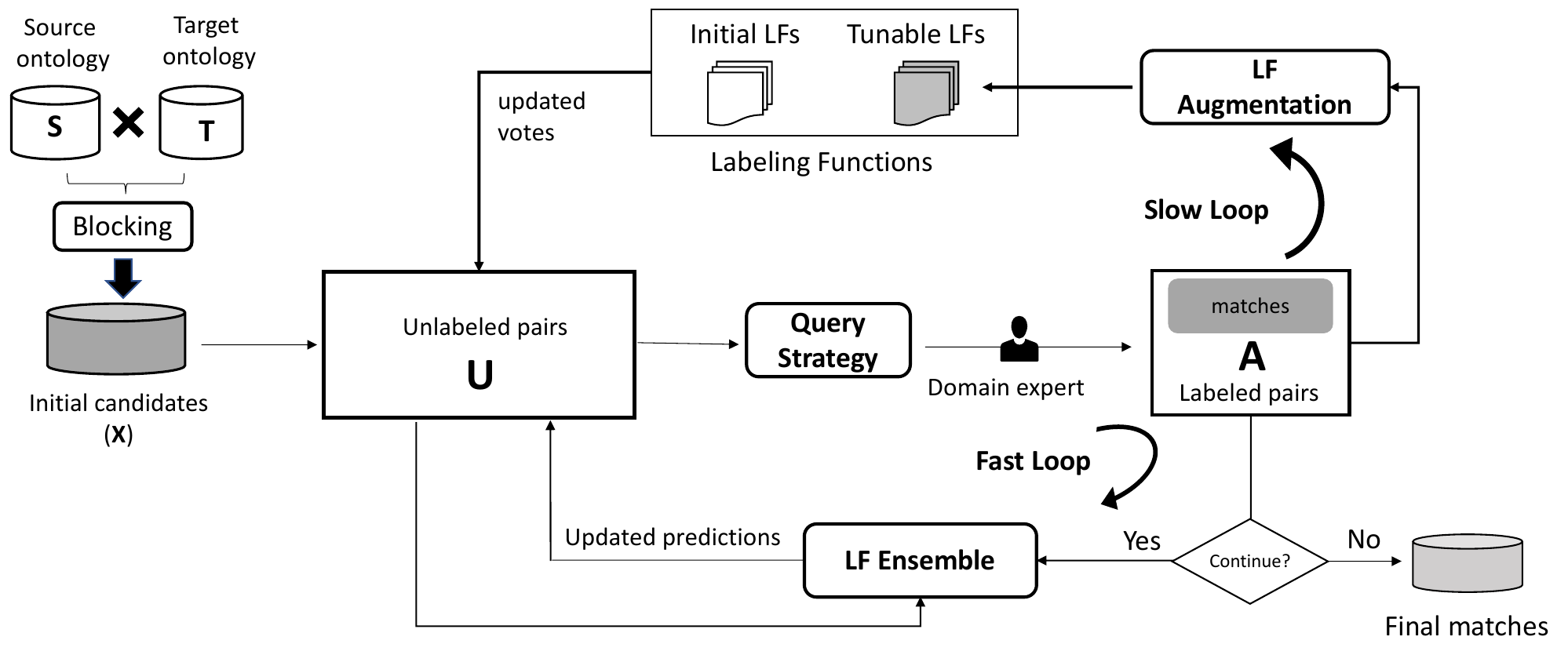}
    \caption{Overview of the DualLoop System. DualLoop encompasses two learning loops that run in parallel: (1) the fast loop picks class pairs, queries the domain expert, and update the prediction results  
for remaining unlabeled class pairs; (2) the slow loop creates and updates tunable labeling functions based on new annotation batches.}
    \label{fig:dualloop_system}
    \rmspace
\end{figure*}

Fig.~\ref{fig:dualloop_system} illustrates the overall design of the DualLoop system. 
As explained in the preceding section, 
DualLoop encompasses two loops that run in parallel but are triggered in a different manner. 
In the fast loop, DualLoop runs its mechanism to pick class pairs, query the domain expert, and update the
prediction results  
for each remaining unlabeled class pair. 
The fast loop is thus triggered by the interactions with the domain expert
and continues until the domain expert decides to stop. 
Once it stops, DualLoop will engage the domain expert 
to verify the predicted matches and then combine them with the already-annotated matches as the final matches. 
In the slow loop, DualLoop creates and updates 
the 
tunable labeling functions.
The slow loop runs 
repeatedly on a parallel processing thread.
Its batch size is se to 2B, twice as large as the batch size of the fast loop.

\subsection{Blocking}
\label{sec:blocking}
Prior to the execution of the two loops,
DualLoop performs 
a key-based blocking process to prepare the initial matching candidates,
filtering out a substantial share of class pairs that are highly unlikely to represent a match. 
Given ontologies $S$ and $T$ with sets of classes $E_S$ and $E_T$, the initial set of potential matching pairs has cardinality $|E_S|\cdot |E_T|$.
The goal of blocking is to eliminate as many candidates as possible without missing any true matches, thus reducing the computational overhead and query cost of the downstream interactive matching phase. In DualLoop we apply a key-based blocking algorithm to prepare the candidates for the interactive learning process. 
It 
selects, for each class from $S$ and each blocking key, the top-k matching partners from~$T$.
We select $k$ as 
\begin{equation}
k =
\begin{cases}
50 & \text { if } |E_T| > 50 \\
|E_T| & \text { otherwise \ .}
\end{cases}
\end{equation}
The union of all resulting candidate pairs then constitutes the final candidate set. We use \texttt{distilbert-base-nli-mean-tokens} based on~\cite{sanh2019distilbert} and \texttt{paraphrase-MiniLM-L6-v2} based on~\cite{reimers-2019-sentence-bert} to create embeddings for computing similarities.
Table~\ref{tab:blocking_key} lists all blocking keys used in our experiments. 

\begin{table}[ht]
\centering
\footnotesize
\caption{List of all blocking keys used by the blocking algorithm; each key is a similarity score of two classes automatically calculated from their attributes, such as class name, label, and comment}
\label{tab:blocking_key}
\begin{tabular}{p{0.45\columnwidth} | p{0.5\columnwidth}}
\hline  
\textbf{key} & \textbf{description}\\
\hline  
\texttt{num\_common\_words} & number of common words among the class name attributes \\ 
\hline 
\texttt{class\_name\_words\_similarity\_a} &  cosine similarity between the class name embeddings computed by sentence transformer \texttt{distilbert-base-nli-mean-tokens}\\
\hline 
\texttt{class\_name\_words\_similarity\_b} & cosine similarity between the class name embeddings computed by sentence transformer \texttt{paraphrase-MiniLM-L6-v2}\\
\hline 
\texttt{label\_words\_similarity\_a} & cosine similarity between the class label embeddings computed by \texttt{distilbert-base-nli-mean-tokens}\\
\hline 
\texttt{label\_words\_similarity\_b} & cosine similarity between the class label embeddings computed by \texttt{paraphrase-MiniLM-L6-v2}\\
\hline 
\texttt{comment\_similarity\_a} & cosine similarity between the comment embeddings computed by \texttt{distilbert-base-nli-mean-tokens}\\
\hline 
\texttt{comment\_similarity\_b} & cosine similarity between the class comment embeddings computed by \texttt{paraphrase-MiniLM-L6-v2}\\
\hline  
\end{tabular}
\end{table}

\subsection{Distance metrics}

In the DualLoop system nine 
fixed distance metrics
are calculated for each matching candidate,
including six embedding distance metrics and three overlap distance metrics. 
The embedding distance metrics are computed from the three given class attributes 
(name, label, comment) using the sentence transformer model~\cite{Huggingface} 
\texttt{distilbert-base-nli-mean-tokens} as well as the transformer model \texttt{paraphrase-MiniLM-L6-v2}.
The overlap distance metrics are calculated only from the name attribute,
using Levenshtein distance, Hamming distance, and the number of overlapping words. 
The nine fixed metrics are additionally used as the input features of four machine learning models (random forest, xgboost, logistical regress, multilayer perceptron) that are repeatedly trained on the set $A$ of annotated data to predict the probability of a given class pair not to be a match. Those four probabilities serve as additional distance metrics.

\subsection{Labeling functions} 

We construct tunable labeling functions from the six embedding distance metrics and the four ML-based metrics.
In DualLoop two types of labeling functions are used in the voting committee: 1)~initial labeling functions that produce static voting results with a simple and easy-to-program logic; 2)~tunable labeling functions that provide dynamic voting results updated based on an automatically tuned threshold. 
The initial labeling functions are more conservative with low coverage, meaning that they 
identify
likely matches with a relatively safe threshold and then leave the rest as undecided,
while tunable labeling functions are more explorative with full coverage,
because their internal thresholds can be automatically tuned with more and more observations from the annotated pairs. 

For the current implementation  15 simple initial labeling functions 
are defined to judge if the two classes in a candidate pair are equivalent based on their common attributes. 
Table~\ref{tab:labeling_functions} describes the details of all initial labeling functions used in DualLoop. 
During the active learning processing,
DualLoop further creates 10 tunable labeling functions. 
6 of them are based on the embedding distance features and the other 4 are based on the probability of trained machine learning models, 
including random forest, xgboost, logistical regress, and 
multilayer perceptron. 

\begin{table}[ht]
\centering
\footnotesize
\caption{List of our 15 initial labeling functions that help to bootstrap DualLoop so it achieves a high sample efficiency.}
\label{tab:labeling_functions}
\begin{tabular}{ p{0.4\columnwidth} | p{0.5\columnwidth}  }
\hline  
\textbf{name} & \textbf{description}\\
\hline  
\texttt{LF\_class\_name\_equal} & check whether the class name attributes are equal \\
\hline 
\texttt{LF\_class\_name\_stemmed\_equal}  & check whether the class name attributes stem from the same words using the \texttt{PorterStemmer} library from NLTK~(\cite{loper2002nltk})\\
\hline 
\texttt{LF\_acronyms} & check whether the acronyms of the class names are equal\\
\hline 
\texttt{LF\_class\_name\_synonyms} & check whether the synonyms of the class names are equal\\
\hline 
\texttt{LF\_label\_equal} & check whether the labels are equal\\
\hline 
\texttt{LF\_root\_nouns\_equal} & check whether the root nouns of the class name attribute are equal \\
\hline 
\texttt{LF\_class\_name\_distance} & check whether the class names are similar based on their distance calculated from fuzz.ratio \\
\hline 
\texttt{LF\_name\_segment\_overlap} & check whether classes are matched based on the intersection of segmented class name attributes\\
\hline 
\texttt{LF\_label\_words\_overlap} & check whether classes are matched based on intersection of segmented label attributes \\
\hline 
\texttt{LF\_subclasses\_overlap} & check whether classes are matched based on overlap of sub-classes\\
\hline 
\texttt{LF\_superclasses\_overlap} & check whether classes are matched based on overlap of super-classes\\
\hline 
\texttt{LF\_properties\_overlap} & check whether classes are matched based on overlap of properties \\
\hline 
\texttt{LF\_class\_name\_similarity} & check whether classes are matched based on embedding distance of class name attributes calculated by 
\texttt{paraphrase-MiniLM-L6-v2}\\
\hline 
\texttt{LF\_label\_similarity} & check whether classes are matched based on embedding distance of label attributes calculated by
\texttt{paraphrase-MiniLM-L6-v2}\\
\hline 
\texttt{LF\_comment\_similarity} & check whether classes are matched based on  embedding distance of  comment attributes calculated by
\texttt{paraphrase-MiniLM-L6-v2}\\ 
\hline  
\end{tabular}
\rmspace
\end{table}  
\section{Experimental Evaluation}

The goal of our experimental design is first to quantify DualLoop's performance gain compared to relevant interactive learning-based and heuristic approaches (Section~\ref{subs:comparision_learning} and~\ref{subs:comparision_heuristic}). We then pay special attention to evaluate recall and query cost efficiency (Section~\ref{subs:recall_query_cost}), followed by an ablation study into DualLoop's core techniques (Section~\ref{subs:ablation}) and runtime efficiency (Section~\ref{subs:runtime_efficiency}).

\subsection{Experimental Setup}

We evaluate the DualLoop approach against the state of the art methods over three datasets and conduct the experiments on one computing server
with 3 GeForce GTX TITAN X GPUs and one Intel Xeon E5-2637 v4 3.50 GHz CPU with 8 cores.
All datasets, code, and configurations for our experiments are shared via 
an anonymous git repository\footnote{https://gitfront.io/r/user-5607906/vmNAegLUYz4q/DualLoop} to reproduce the evaluation results presented below. 

\subsection{Datasets}

We use three open datasets from different domains with various sizes.
The first dataset is from the conference track of the Ontology Alignment Evaluation Initiative~\cite{OAEI-Conference}, including 7 different ontologies for the domain of scientific conferences, leading to 21 ontology pairs (\emph{matching tasks}) in total. The ground truth data of those tasks is provided by OAEI. 
The second dataset is from an open challenge called Knowledge Extraction for the Web of Things (KE4WoT)~\cite{KE4WoT}. We use four ontologies related to the Internet of Things and manually create the ground truth for three matching tasks. 
The third and final dataset contains a single matching task, consisting of two widely used ontologies that were
independently developed in the U.S. and Europe for the air traffic management domain~\cite{vennesland2019matching}. This dataset has been shown to be very challenging for ontology matchers~\cite{furst2023versamatch}.
Its ground truth data was provided in a previous study~\cite{atmonto2airm}. 
Table~\ref{tab:dataset} summarizes the statistical features of 
these three datasets after application of the 
blocking step.
In addition, we list the key parameters used in our experiments in the table. 
In each experiment, the batch size in the slow loop is twice of the one in the fast loop. 

\begin{table}[!th]
\centering
\caption{Summary of datasets and parameters after application of the blocking step.}
\label{tab:dataset}
\begin{tabular}{ l | l | l | l  }
\hline
\textbf{dataset name} & \textbf{Conference} & \textbf{AI4EU} &\textbf{AirTraffic}\\
\hline
matching tasks & 21 &  3 &  1 \\
\hline
classes in S & 29 - 140 & 92 &  154 \\
classes in T & 38 - 140 & 451 -1,035 & 915 \\
\hline
avg. candidates & 2,376 & 8,214 & 12,213 \\
\hline
true matches & 4 - 23  & 17 - 33 &  32 \\
avg. percentage & 0.60\%  & 0.30\% & 0.26\% \\ 
\hline  
batch size ($B$) & 10 & 50 & 100\\
increment ($\delta$) & 0.02 & 0.02 & 0.02\\
\hline
\end{tabular}
\rmspace
\end{table}

The overall performance of DualLoop is evaluated based on the comparison against two types of approaches to interactive ontology matching: 
1) \emph{learning-based approaches} that are able to make query decisions based on the newly updated estimation given by a machine learning model retrained after each user annotation round;
2) \emph{heuristic-based approaches} that are designed to filter out the noise in the selected mappings for better precision
using some rule-based reasoning and inconsistency checking. 
The detailed results are presented below. 

\begin{figure*}[!th]
\centering
\subfloat[AI4EU]{%
\includegraphics[width=0.35\linewidth]{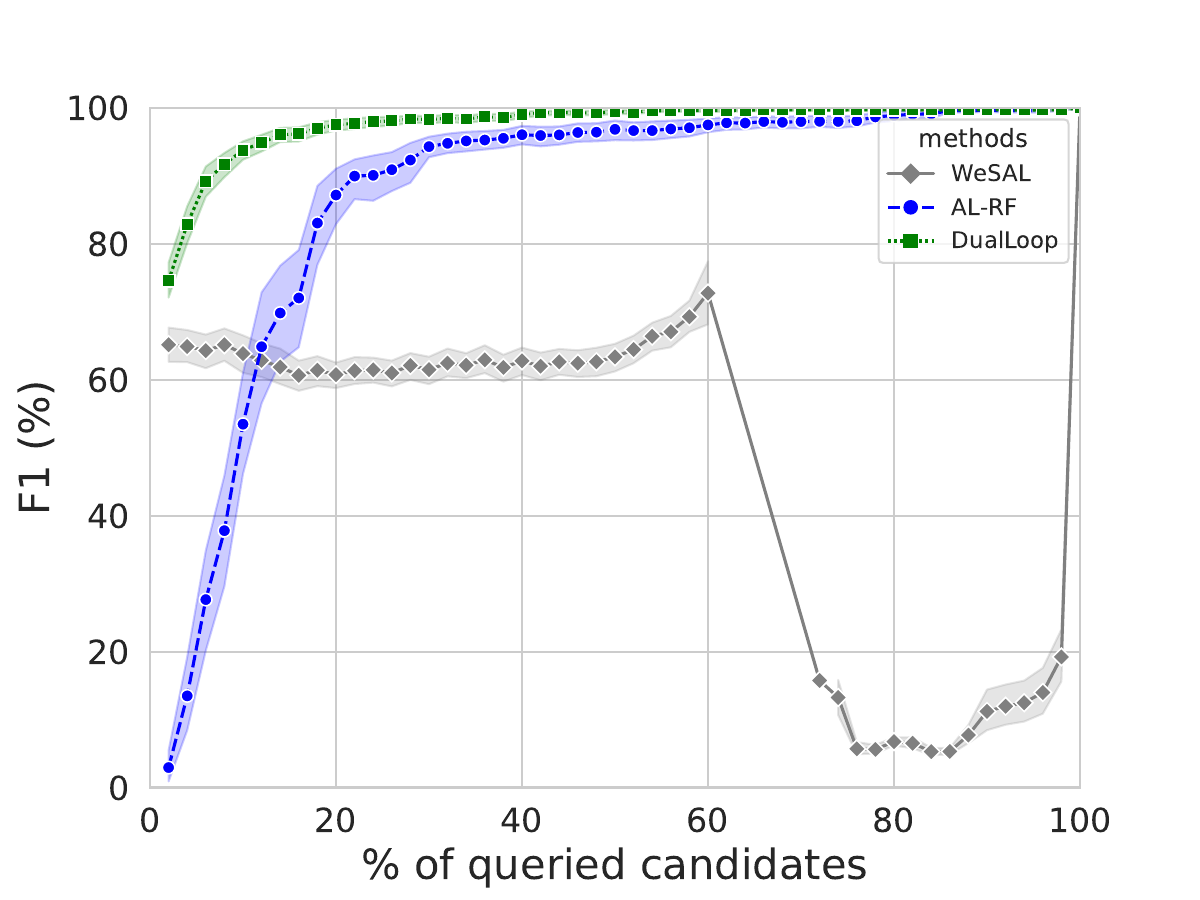}%
\label{fig:f1_conference}%
}%
\hspace{-2em}
\subfloat[AI4EU]{%
\includegraphics[width=0.35\linewidth]{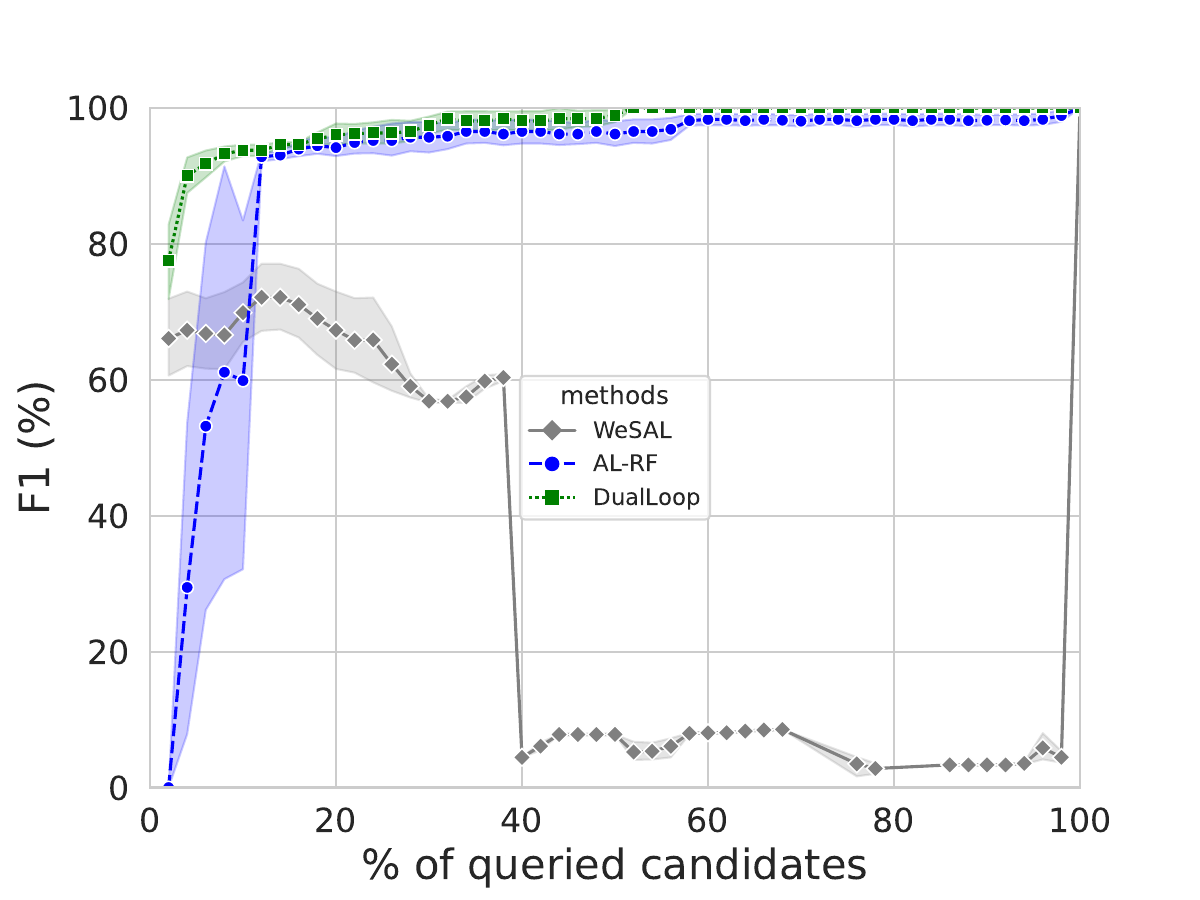}%
\label{fig:f1_ai4eu}%
}%
\hspace{-2em}
\subfloat[AirTraffic]{%
\includegraphics[width=0.35\linewidth]{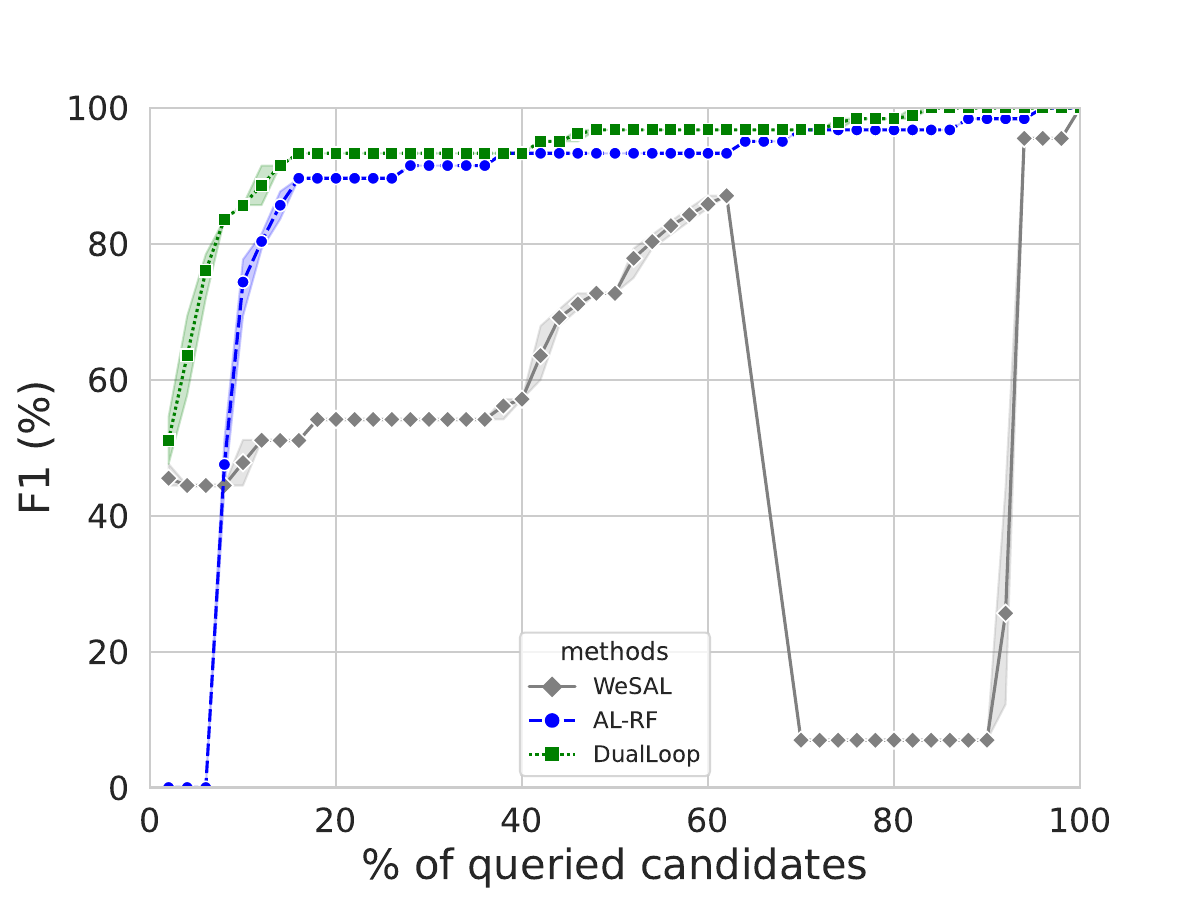}%
\label{fig:f1_nasa}%
}%
\caption{F1 score plotted against the percentage of overall query budget for different datasets. DualLoop clearly achieves higher F1 scores than the other two methods on all datasets by improving the matching quality in face of small query budgets.}
\label{fig:f1_result}
\end{figure*}

\begin{figure}
    \centering
    \includegraphics[width=1.0\linewidth]{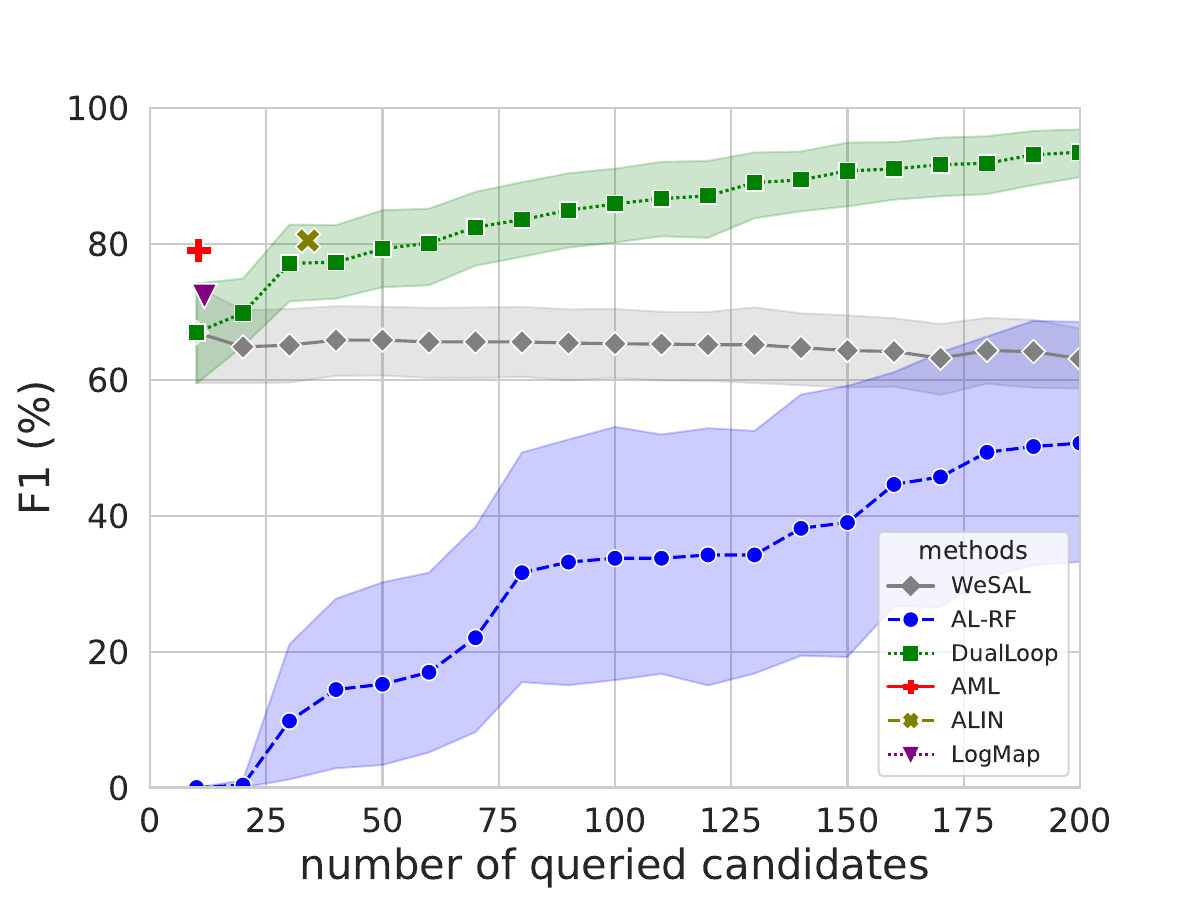}
    \caption{F1 score achieved on the conference dataset~\cite{OAEI-Conference} by our approach (DualLoop), as compared to existing active learning methods without (AL-RF) and with weak supervision (WeSAL), plus state of the art interactive ontology matchers (AML and LogMap). DualLoop already outperforms all other methods with less than 30 user annotations.}
    \label{fig:f1_conference_with_aml}
    \rmspace
\end{figure}

\subsection{Comparison with Other Learning-based Approaches}
\label{subs:comparision_learning}

We evaluate against the following two learning-based methods 
that are most relevant to our approach.  

\begin{itemize}
    \item AL-RF, a conventional active learner. This approach utilizes a single active learning loop in which a random forest model is repeatedly trained based on the data points annotated so far. Its sample query strategy is to maximize the entropy-based uncertainty.
    \item WeSAL, an active learner based on weak supervision. This approach also has a single active learning loop in which an ensemble model is trained based on all data points, using a fixed set of labeling functions. Its sample query strategy is to maximize the disagreement of votes given by all labeling functions. We use the same set of labeling functions as in DualLoop.
\end{itemize}

The two baselines are selected based on three considerations.
First, they cover representative state-of-the-art methods, 
including one method for conventional active learning 
and one
for active learning combined with weak supervision. 
Second, random forest was demonstrated to be the most efficient active learning model for related tasks such as entity matching. 
Last, WeSAL is the method closest to DualLoop in terms of combining active learning and weak supervision with sample-based expert feedback.

We experimentally analyze the matching performance of DualLoop and the baselines in terms of F1 score for varying numbers of oracle queries. Subsequently, we compare the total query cost with the total amount of matches found, which includes both the active learning phase and a final verification of the identified queries by the domain expert.

\paragraph{Analysis of F1 score} 
We examine how matches and non-matches in the entire candidate set
can be precisely identified as further data points are annotated by the domain expert. 
We use F1 score 
to quantify the performance with regard to the trade-off between precision and recall. 

For each matching task, the maximum query budget is set to the number of matching candidates.
Figure~\ref{fig:f1_conference}),~\ref{fig:f1_ai4eu}), and~\ref{fig:f1_nasa}) show 
how the F1 scores 
change 
with the percentage of the query budget used. 
In the figures, the points represent the mean F1 score achieved by the three evaluated methods across the matching tasks in the respective data set, plotted in steps of 2\% of the maximum budget. The lines represent linear interpolation, while the shaded areas represent the confidence region of 95\%~\cite{ConfidenceInterval}.

DualLoop clearly achieves higher F1 scores than the other two methods on all datasets.
It first bootstraps the initial F1 score similarly to WeSAL, and then improves much faster than AL-RF at the early active learning stage. 
For example, in the AirTraffic dataset, 
with a 10\% query budget DualLoop achieves F1 score of 80\% while 
both WeSAL and AL-RF are below 60\%.
Notice that AL-RF lags behind at the beginning of the process due to the lack of positive samples. 
This demonstrates that, with highly imbalanced data, application of random sampling without prior knowledge is a considerable disadvantage.
However, AL-RF quickly catches up once a few true matches are found. In contrast, WeSAL starts with a reasonable F1 score 
due to the help of its labeling functions, 
but benefits much slower from annotated data points. 
In some case its F1 score even decreases with more annotations. This can be explained with low precision for the remaining data points that are not covered by the initial labeling functions. Thus, WeSAL falls behind AL-RF after a certain number of data points have been annotated.  
Overall, the evaluation demonstrates that DualLoop overcomes the limitations of both AL-RF and WeSAL across all three datasets and especially improves the matching quality in face of small query budgets.

\begin{figure*}[!th]
\centering
    \subfloat[Conference]{%
            \includegraphics[width=0.35\linewidth]{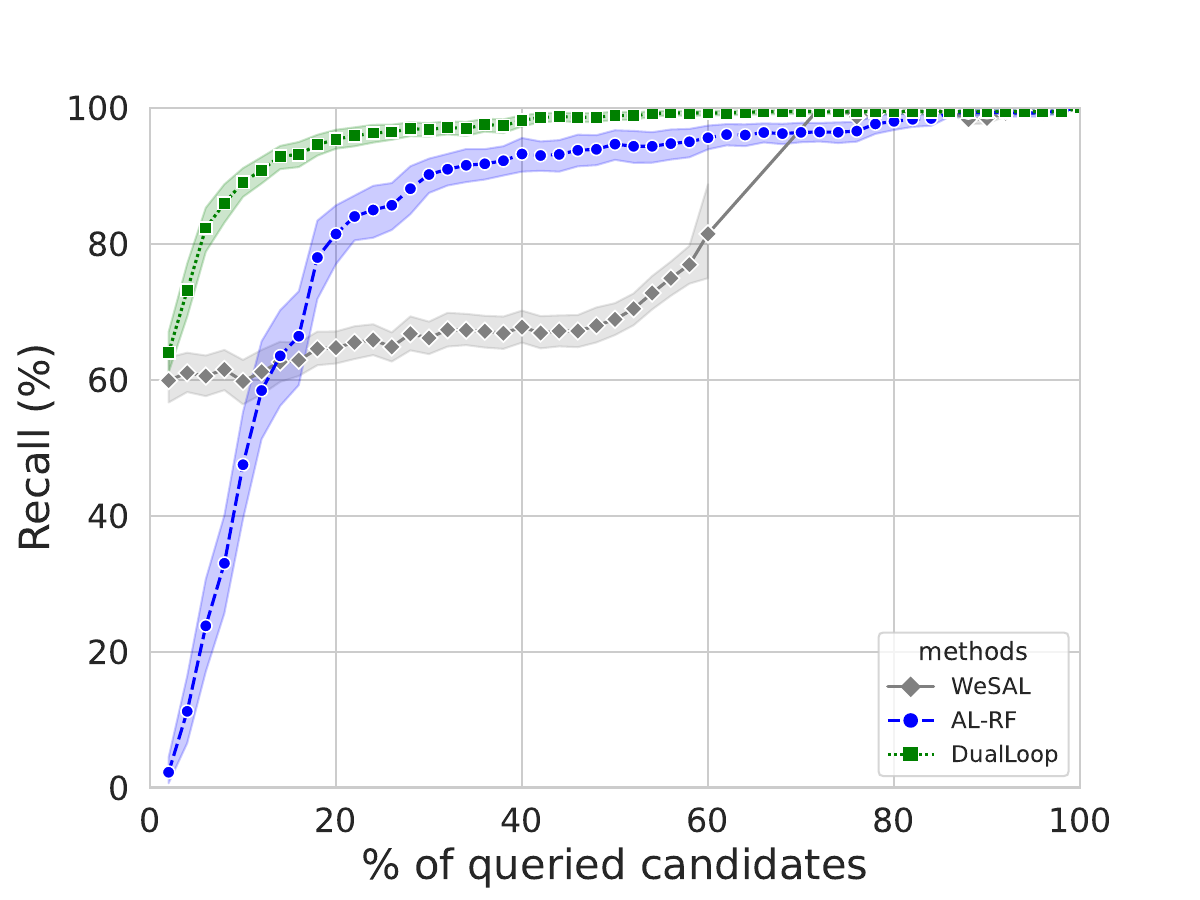}
            \label{fig:recall_conference}
    }
    \hspace{-2em}
    \subfloat[AI4EU]{%
            \includegraphics[width=0.35\linewidth]{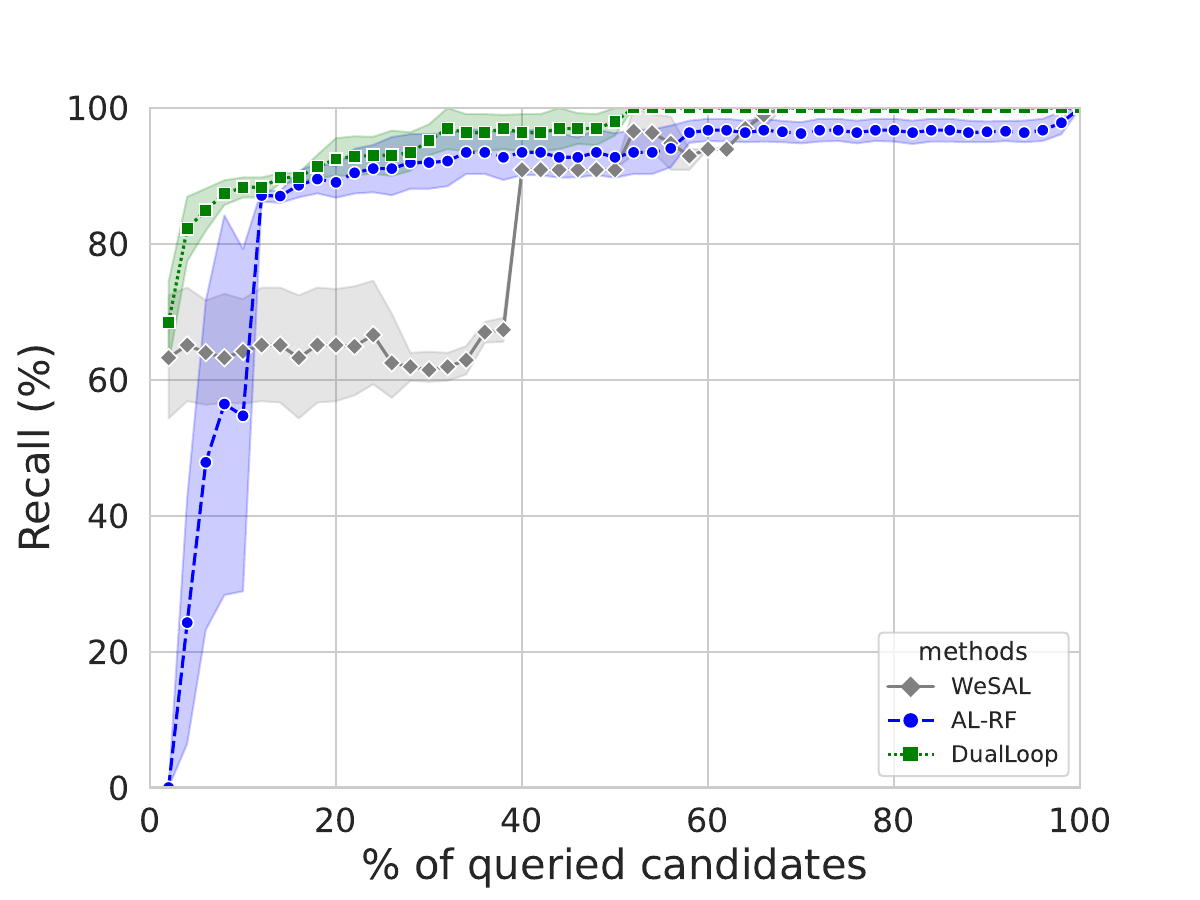}
            \label{fig:recall_ai4eu}
    }
    \hspace{-2em}
    \subfloat[AirTraffic]{
            \includegraphics[width=0.35\linewidth]{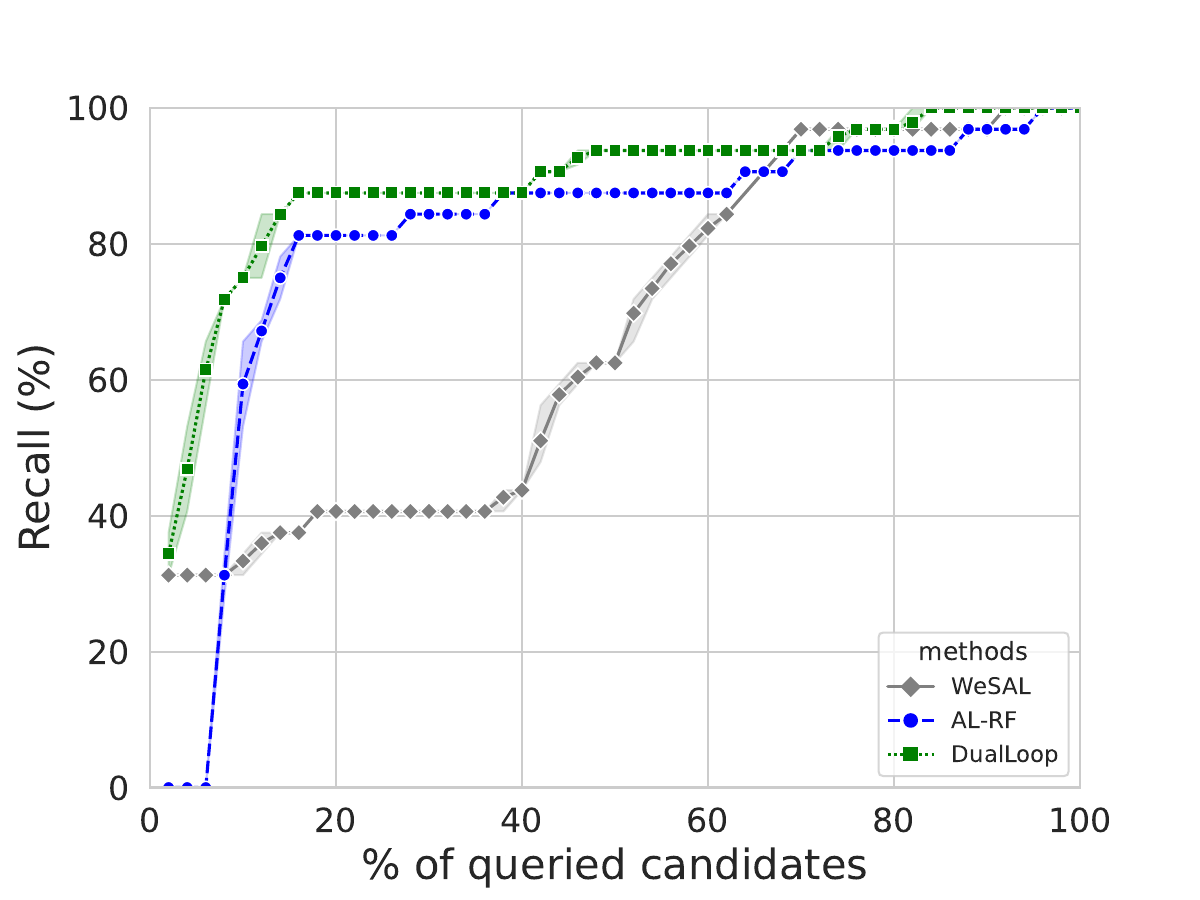}
            \label{fig:recall_nasa}%
    }    
\caption{Recall plotted against query budget including match verification. DualLoop achieves substantially higher recall on all three datasets
in the early stage and maintains a competitive recall for the entire query budget range.  }
\label{fig:recall_result}
\rmspace
\end{figure*}

\subsection{Comparison with Heuristic Approaches}
\label{subs:comparision_heuristic}

We also compare DualLoop against three existing interactive ontology matching methods that are not based on active learning, namely 
AML~\cite{faria2013agreementmakerlight}, ALIN~\cite{ALIN2020}, and LogMap~\cite{jimenez2011logmap}.
Figure ~\ref{fig:f1_conference_with_aml}) shows the average F1 score per absolute number of queried candidates 
across all 21 matching tasks in the Conference dataset. The results of AML, ALIN, and LogMap are plotted as single points 
because those approaches automatically decide how many candidates to query. For comparison, we take their best results reported by OAEI in 2021~\cite{OAEI-AML}. The number of queried candidates for them is calculated
by dividing the total number of distinct mappings with zero error rate by 21 (the number of matching tasks). 
Although they show slightly better F1 score than DualLoop with less than 30 queried candidates, DualLoop already outperforms them 
after having learned from 50 queries. 
Furthermore, by querying 200 candidates,
\emph{DualLoop reaches an F1 score of 96\%, 
which has never been reported for any method in the OAEI community. }

\begin{table}[!t]
\centering
\caption{Query cost required to achieve a specific recall. DualLoop achieves the same recall levels as other methods with only a fraction of annotation cost.}
\label{tab:cost_reduction}
\begin{tabular}{p{0.15\linewidth}p{0.14\linewidth}rrrr}
\hline
\multirow{2}*{\textbf{Dataset}} & \multirow{2}*{\textbf{method}} & \multicolumn{4}{c}{\textbf{recall (\%)}}\\
 & &\textbf{70}&\textbf{80} &\textbf{90}&\textbf{98}\\
\hline
\multirow{3}*{Conference}  & WeSAL & 22,329 & 34,103 & 37,919 & 39,103\\ 
 & AL-RF & 8,852 & 11,051 & 15,731 & 20,191\\ 
 & DualLoop & \textbf{1,892} & \textbf{2,870} & \textbf{6,080} & \textbf{13,290}\\ 
\hline
\multirow{3}*{AI4EU} & WeSAL & 7,894 & 7,894 & 11,425 & 13,844\\ 
 & AL-RF & 1,652 & 1,802 & 10,000 & 17,100\\ 
 & DualLoop & \textbf{390} & \textbf{886} &	\textbf{3,600} & \textbf{8,350}\\  
\hline
\multirow{3}*{AirTraffic}  & WeSAL & 6,213 & 7,013 & 8,441 & 11,041\\ 
 & AL-RF & 1,500 & 1,800 & 7,600 & 11,500\\ 
 & DualLoop & \textbf{800} & \textbf{1,400} & \textbf{4,900} & \textbf{10,000}\\ 
\hline
Avg. cost & WeSAL & 91.2\% & 86.8\% & 64.8\% & 38.4\% \\
reduction & AL-RF & 67.2\% & 49.0\% & 53.6\% & 32.8\% \\
\hline
\end{tabular}
\vspace{-2ex}
\end{table}

\subsection{Recall and Query Cost}
\label{subs:recall_query_cost}
Here we evaluate DualLoop and the baselines under the assumption that all matches need verification by a domain expert, a scenario which is motivated by the asymmetrical misprediction cost explained in Section~2. In this setting, false positives increase the number of queries. 
False negatives have a negative effect on the recall, which
is calculated as the ratio of true matches that are identified either during the active learning phase or during verification of the predicted matches. Likewise, the query costs are calculated as the number of annotated samples plus the number of matches to verify.

Figure~\ref{fig:recall_conference}),~\ref{fig:recall_ai4eu}), and~\ref{fig:recall_nasa}) 
visualize how recall changes 
as the query cost increases from 0 to 100\% of the matching candidates. 
We see a similar pattern across all datasets.
DualLoop and WeSAL start with the same recall as they initially share the same set of labeling functions. 
However, within a few queries, DualLoop can already bootstrap its learning capability to find more matches than the other two methods
with the same query cost, 
thanks to the additional labeling functions with their tuned thresholds. 
AL-RF achieves a large recall increase once it has collected a reasonable number of positive and negative annotations, 
but with a query budget less than 5\% it has a recall of
less than 20\% in most cases, which is far below what DualLoop achieves. 
With a medium query cost, from 10\% to 40\%, 
AL-RF catches up and often overtakes WeSAL,
but both have lower recall than DualLoop. 
Overall, DualLoop achieves substantially higher recall 
on all three datasets
in the early stage 
and maintains a competitive recall
for the entire query budget range.  

Table~\ref{tab:cost_reduction} further displays results from the perspective of cost reduction, listing the absolute query cost that each method spends for achieving a given recall, ranging from 70\% to 98\%. 
For the datasets with multiple matching tasks the query cost is averaged over all tasks. 
DualLoop requires less queries to achieve the same recall. For example, the average cost for reaching 90\% recall is reduced by 64.8\% and 53.6\% as compared to WeSAL and AL-RF, respectively.

\subsection{Ablation Analysis}
\label{subs:ablation}
\begin{figure}[!t]
\centering
    \includegraphics[width=1.0\linewidth]{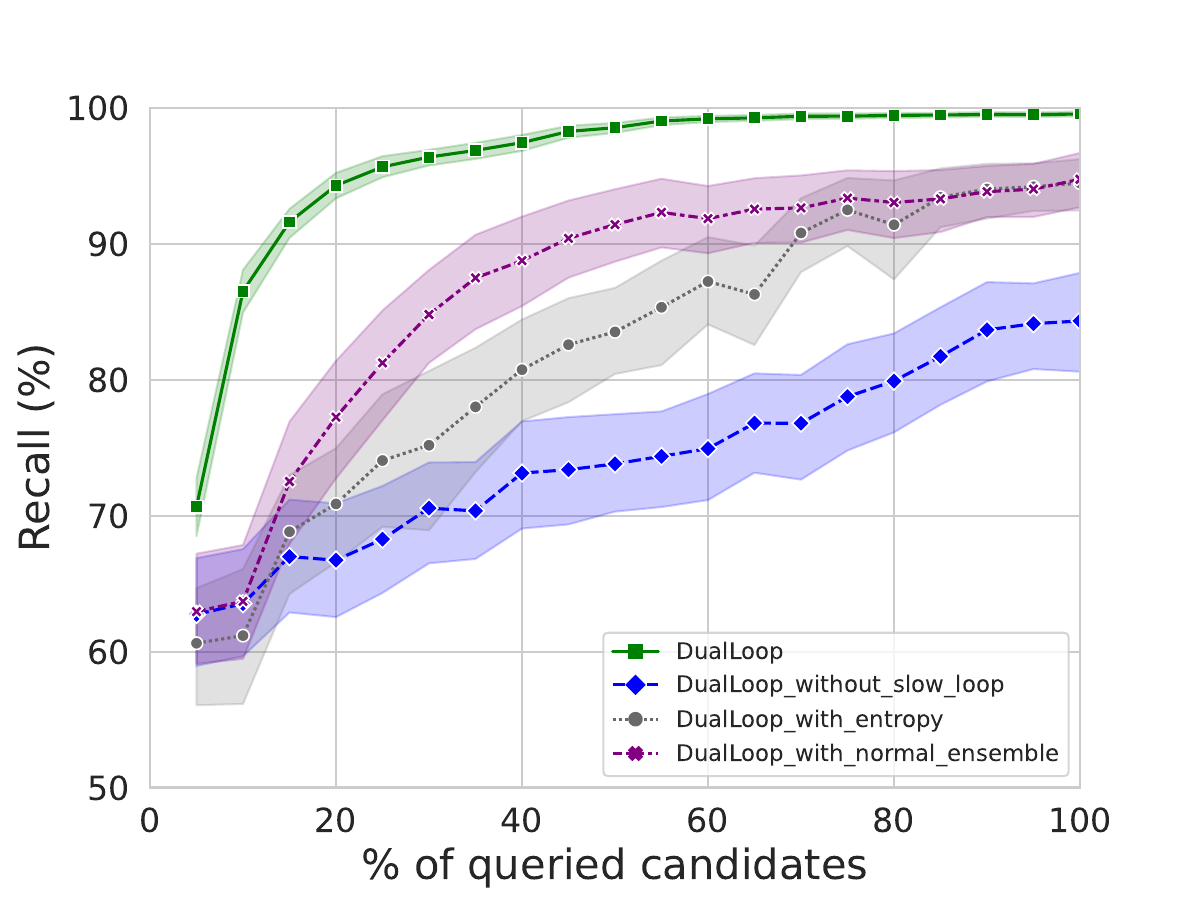}
    \label{fig:recall_breakdown}
\caption{Results of various settings for all datasets. All three techniques make a significant contribution to the overall performance of DualLoop.}
\label{fig:ablation_study}
\rmspace
\end{figure}

We study the contribution of different ingredients within DualLoop. 
By deactivating or replacing only one ingredient each time, we compare three cases with the full DualLoop method: 
(1) \emph{DualLoop\_without\_slow\_loop}, where the slow loop is deactivated, so that the labeling functions within the voting committee are static,
(2) \emph{DualLoop\_with\_entropy}, which replaces the DualLoop sample query strategy with the conventional entropy-based strategy, 
and (3) \emph{DualLoop\_with\_normal\_ensemble}, which replaces the DualLoop ensemble with the one used by Snorkel to disable the selection and weighting of labeling functions based on the query results.
Figure~\ref{fig:ablation_study} displays the comparison
in terms of mean recall, aggregated over all three datasets and tasks, as well as the 
95 percent confidence interval. 
We observe that \emph{all three ingredients make a significant contribution to the overall performance of DualLoop}. 
The largest performance drop happens when the slow loop is deactivated, indicating the clear benefit of running both the slow loop with long-term learners and the fast loop with the short-term learner, which is the key feature of DualLoop.

\subsection{Runtime Efficiency}
\label{subs:runtime_efficiency}
Last, we profile the computation time of each step in DualLoop to 
analyze how the separation of slow and fast loop helps to reduce 
the response time of user interaction and increase the throughput of expert queries.
The largest share of computational costs are due to the pre-processing step to
create the embedding features of each class, which is executed only once 
before the start of the interactive learning part.
The two steps (sample query and labeling function ensemble) of the fast loop 
run indeed very fast, with a latency of 0.25 seconds between subsequent queries on average.
The ensemble augmentation step in the slow loop needs on average 12 seconds per iteration, 
because it involves training of machine learning models as well as parameter searching. 
When more advanced models like BERTMap~\cite{he2022bertmap} are utilized in the slow loop, 
the latency gap between the fast loop and the slow loop could be even larger. 
This gap justifies the benefit of the two-loop approach in DualLoop,
which sheds light on the design of future interactive machine learning systems
to balance both learning capability and responsiveness.  
\section{Usage in Commercial Cases}

The core method of the DualLoop system has been integrated into a commercial toolkit called \emph{TrioNet} to provide interactive ontology matching service in various business domains. 
This section introduces the system overview of TrioNet and reports its performance results and the comparison analysis against other existing tools 
in context of an executed project in the Architectural Engineering and Construction (AEC) sector.

\subsection{TrioNet Toolkit}
\label{subs:trionet}

\begin{figure}[ht]
    \centering
    \includegraphics[width=1\columnwidth]{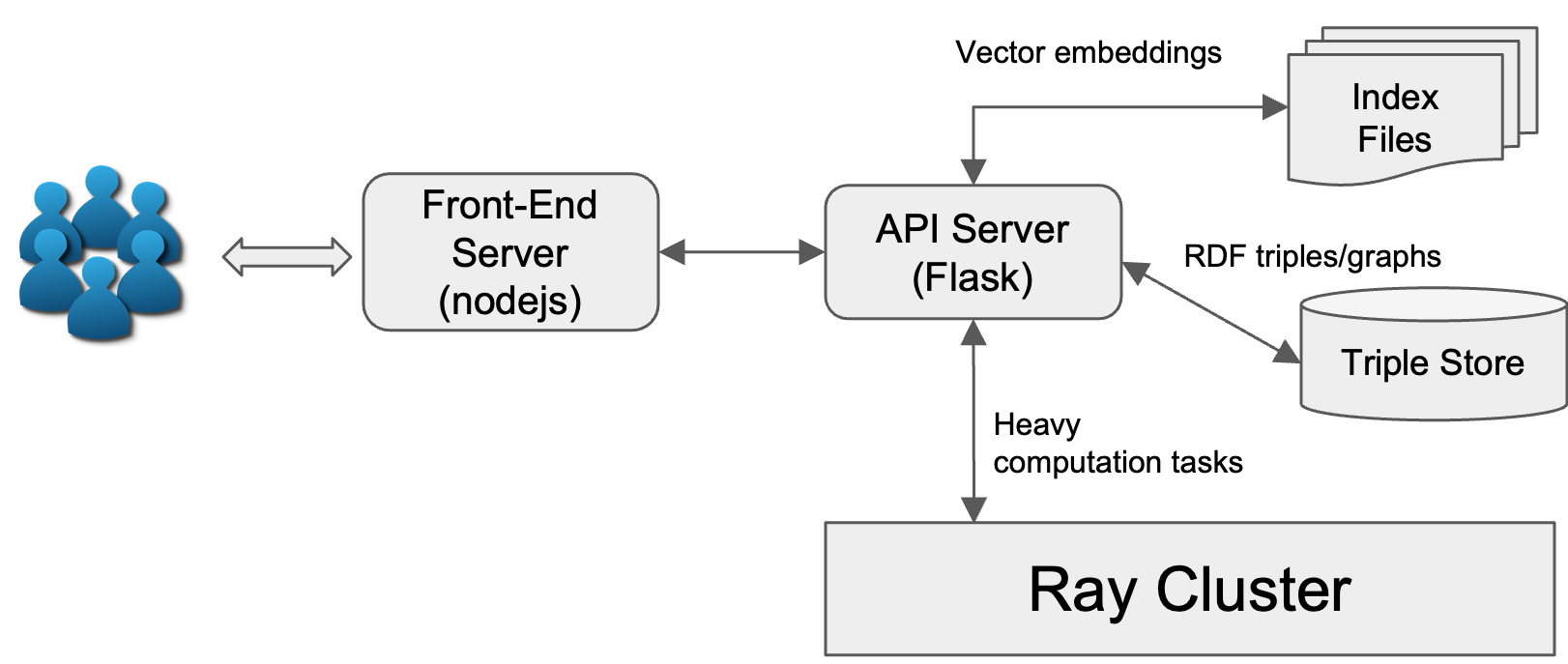}
    \caption{Overview of the TrioNet Toolkit that implements the Dualloop method. Users interact with the system through a graphical web interface, while Faiss~\cite{douze2024faiss} and Ray~\cite{moritz2018ray} speed up similarity search and embedding generation respectively to make the toolkit responsive.}
    \label{fig:trionet}
    \rmspace
\end{figure}

The TrioNet toolkit, depicted in Figure~\ref{fig:trionet}, comprises five major system components:

\begin{enumerate}

\item The front-end server, developed in JavaScript using Node.js, facilitates web-based interfaces for users to manage ontologies and verify matches suggested by DualLoop.

\item The API server, implemented in Python using Flask, serves REST requests from the front-end server, executing core data management logic and learning algorithms.

\item The triple store, powered by Virtuoso~\cite{erling2012virtuoso}, manages all imported ontologies and provides SPARQL queries to the API servers.

\item A collection of index files, generated using the Faiss library~\cite{douze2024faiss}, serves to backup and restore embedding vectors for each concept within each ontology.

\item A Ray~\cite{moritz2018ray} cluster, comprising multiple computer nodes, facilitates parallelized embedding generation from various pretrained large language models.

\end{enumerate}

This integrated toolkit forms the backbone of TrioNet, enabling efficient ontology management, matching, and learning functionalities.

\subsection{TrioNet Functionalities}

TrioNet offers three primary functionalities:

\begin{itemize}
    \item \textit{Ontology Management}: Users can upload and remove ontologies and explore concepts within a selected ontology through an interactive graphical user interface. Upon uploading a new ontology, TrioNet initiates the generation of embedding vectors for each concept within the ontology. These vectors are subsequently utilized for concept search and ontology matching.
    
    \item \textit{Concept Search}: TrioNet facilitates finding top matches for any given concept, whether it's a newly input term or an existing concept from an uploaded ontology, using the blocking algorithm of the DualLoop system.
    
    \item \textit{Ontology Matching}: This constitutes the core service of TrioNet, empowered by DualLoop, for identifying matched concepts between two selected ontologies. Leveraging the cost-efficient active learning mechanism of DualLoop, TrioNet allows users to efficiently narrow down potential matches by verifying concept pairs suggested by DualLoop. During each verification step, DualLoop selects 10 candidate pairs, presenting them within a table alongside descriptions and predicted matching results for user verification. Users can either confirm predictions directly or revise them. Subsequently, identified matches are utilized to create 'sameAs' triples, linking concepts between the source and target ontologies. These triples are maintained in a separate graph for querying alongside other ontology graphs. Through this iterative verification process, TrioNet enables users to discover more matches with the same or lower verification effort compared to existing approaches.
\end{itemize}

\subsection{Interlinking Ontologies in the AEC Industry}

TrioNet has been instrumental in several of our commercial solutions for interlinking available ontologies to facilitate data integration. This section presents our experience and evaluation results of employing TrioNet in an industrial use case to identify matched concepts across multiple ontologies within the Architecture, Engineering, and Construction (AEC) industry sector.

Traditionally, Information Foundation Classes (IFC) have served as the reference ontology in the AEC domain, providing semantic descriptions and relationships for building-related assets. However, the Brick community and their ontology-based schema~\cite{balaji2016brick} have recently standardized broader representations of buildings into an integrated cross-domain data model. This effort marks one of the first successful attempts to model transversal elements such as sensor or actuator subsystems within buildings. Consequently, we chose the Brick ontology as the target ontology for a digital building twin project. Given that existing subsystems and domains (e.g., IoT, architecture) often employ their own data models, a semi-automated ontology matching mechanism is essential to ensure reliable interlinking of cross-domain ontologies with minimal human effort in practice.

Numerous alignment efforts with Brick have emerged in recent years, most of which have relied on manual alignments. For instance, Balaji \textit{et al.}~\cite{balaji2016brick} identified 13 elements from the IFC with a semantic match to classes in the Brick ontology. Despite presenting an automatic tool to merge IFC with Brick, only 5 mappings were identified out of the 1,807 elements in the version of the IFC model used. In~\cite{lange2018evaluation}, the authors identified 57 elements from the IFC model that could populate the Brick ontology, while~\cite{luo2022extending} proposed an extension of Brick to model contextual, behavioral, and demographic information linked to building occupants.

In our work, we select a set of commonly found ontologies in the AEC sector and align them with Brick. Specifically, we match the following ontologies against Brick to create an interlinked ontology (accessible in~\cite{garrido2022interlinking}, which we use in this work to evaluate quantitatively the effectiveness of TrioNet implementation in practice). 

\begin{itemize}
    \item \textit{IFC} (ifcOWL v4.0.2.1.). IFC model representation is based on semantic web and linked data technologies for data management and exchange in the AEC industry.
    \item \textit{SAREF} (v3.1.1.). This ontology provides semantic modelling of various activity sectors in the IoT and smart data spaces domains. 
    \item \textit{SAREF4BLDG} (v1.1.2.). This ontology is an extension of SAREF covering both AEC and Internet of Things (IoT) domains.
    \item \textit{IoT-Lite} (rev. 11-26-2015). Lightweight ontology based on IoT resources and services.
    \item \textit{BOT} (rev. 6-28-2021). This ontology provides a high-level description of the topological elements of a building, including 3D assets.
\end{itemize}

We have chosen this industrial sector to practically validate the TrioNet toolkit and to examine the primary challenges encountered in aligning different ontologies, comparing TrioNet with state-of-the-art matchers like AML or LogMap. The insights and lessons learned from this practical experience serve to confirm the suitability of the TrioNet toolkit for industrial applications.
Table~\ref{tab:total_mappings} provides a summary of (i) the total number of classes included in each ontology, and (ii) the number of alignments with Brick in the ground truth obtained through manual annotation.

\begin{table}[htb]
\caption{Number of classes of used ontologies and manually-verified alignments (matches) with Brick~\cite{balaji2016brick}, our target ontology, which contains 1440 classes.}
\label{tab:total_mappings}
\resizebox{0.8 \columnwidth}{!}{
\begin{tabular}{l|c|c}
\hline
\textbf{Source} & \multicolumn{1}{l|}{\textbf{Classes}} & \multicolumn{1}{l}{\textbf{Alignments with Brick}} \\ \hline
IFC & 1285 & 22 \\ \hline
SAREF & 81 & 13 \\ \hline
SAREF4BLDG & 71 & 21 \\ \hline
IoT-Lite & 20 & 3 \\ \hline
BOT & 8 & 6 \\ \hline
\end{tabular}
}
\rmspace
\end{table}

\begin{figure}[ht]
    \centering
    \includegraphics[width=1\columnwidth]{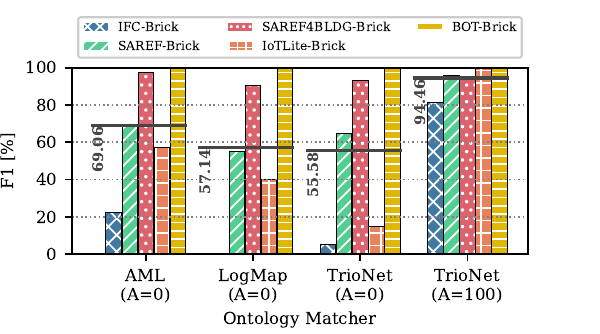}
    \caption{F1 scores achieved by AML, LogMap and TrioNet for different ontology-matching tasks. With only 100 user annotations, TrioNet achieves a $>25$ point higher F1 score across all datasets.}
    \label{fig:bar_plot_trionet}
    \rmspace
\end{figure}

\textbf{Results.} 
Fig.~\ref{fig:bar_plot_trionet} displays an overall comparison of the performance of the three ontology matchers used in the experiments based on the F1 score. TrioNet was evaluated with and without annotations to demonstrate the impact of the active learning loop on matching real ontologies from the AEC sector. 

While F1 scores achieved by TrioNet without using annotations are slightly lower than that of other heuristic-based ontology matchers ---55.58\% on average, compared to 69.06\% (AML) and 57.14\% (LogMap)--- \emph{adding the active-learning loop resulted in an F1 score of 94.46\%. This represents 36.77\% improvements compared to AML and 71.31\% compared to LogMap.}

The overall verification effort is measured by the sum of the number of annotations for checking the suggested concept pairs during the verification steps and the number of annotations for checking all predicted matches given by the final prediction step. 
Although relatively large ontology models require high verification effort of examining predicted matches, a few rounds of verification can significantly reduce the required effort. 
For instance, when matching IFC with Brick via TrioNet, 
incorporating only 40 verifications after the initial four rounds of verification
could reduce the overall verification effort from 730 to only 53 examinations . 

\subsection{Discussion}

{\bf Heuristic Rules vs. Active Learning: } Existing ontology matching tools like AML and LogMap are implemented with a set of sophisticated heuristic rules. They attempt to find potential matches through their best one-shot effort. However, this one-shot strategy tends to make them very conservative, resulting in providing only a set of safe matches. The downside of this approach is that they often identify only the same set of easy matches and do not assist in finding challenging matches. In contrast to those tools, TrioNet can iteratively enhance its predictions through a carefully designed active learning process. While it may not initially achieve the highest cost-efficiency in terms of F1, it can rapidly improve its cost-efficiency after a few verification iterations and ultimately discover more matches. \emph{This has been a well appreciated feature by our in-house customers.}


{\bf When is it Good Enough to Stop: } TrioNet allows users to improve the final prediction with more verified matches or non-matches. One of the natural questions is when it is good enough to stop verification. A good practice is to carry out a few rounds of verifications at the beginning so that TrioNet can estimate the performance of its labeling functions from a representative sample. After that, a good indicator to watch for is the total of verified matches examined during the verification phase and predicted matches estimated by TrioNet after each verification step. \emph{If this total remains unchanged in recent verification steps, it indicates that TrioNet has reached a stable state for making the best prediction.}

{\bf What if Users Make Errors:} another concern is that users might make mistakes by providing incorrect verifications. In practice, it is hard to completely avoid this, but to alleviate this problem, TrioNet introduces an observation list for users to keep track of all concept pairs that they are not sure about during the verification process. In addition, the observation list is shared across users so that those unsure concept pairs can be double checked together with other domain experts. \emph{In a wider deployment automated error detection techniques from crowdsourcing methods could be applied~\cite{hung2017answer}.} 
\section{Related Work}
We first discuss related works from a problem domain perspective in Section~6.1 and then from a method perspective in Section~6.2. 

\subsection{Interactive ontology and entity matching}
Several interactive ontology matchers~\cite{interactiveOM-2021} have been proposed.
They often formulate ontology matching 
as a constraint-based optimization problem 
and solve it with a minimal number of user queries.   
For example, three ontology matchers, namely AML~\cite{faria2013agreementmakerlight}, LogMap~\cite{jimenez2011logmap}, and ALIN~\cite{ALIN2020}, 
have been evaluated for the Interactive Track of Ontology Alignment Evaluation Initiative (OAEI) in the past years. 
While those approaches are query-efficient, they lack the flexibility of active learning based methods to adapt the query budget to the application needs and availability of domain experts.
In addition, a few studies on interactive entity matching or entity resolution have proposed methods to identify duplicated entities within the same table or matched entities between tables, 
through either crowd-sourcing or active learning~\cite{gokhale2014corleone}
~\cite{das2017falcon}
~\cite{firmani2016online}
~\cite{vesdapunt2014crowdsourcing}. 
All of them consider user feedback to improve the matching results, 
but they are not designed to handle highly imbalanced data and their setup does not consider costs of verifying the predicted matching results, which is unsafe in light of the high cost of false positives.
To the best of our knowledge, our work is the first active learning-based approach to ontology matching that successfully overcomes the problem of imbalanced data, and our study is the first to consider the human involvement cost for both the interactive phase and the verification of matching results.

\subsection{Weak supervision with active learning}
Traditional active learning approaches target to learn an accurate predictor 
based on samples selected by the algorithm and annotated by domain experts. 
To select maximally informative samples to query, many rely on uncertainty sampling~\cite{Abdar2021}, but they face the problem to bootstrap the query process at the beginning.
To avoid this problem, weak supervision~\cite{ratner2020snorkel} has been explored together with active learning 
to initially utilize weak labels computed by a set of user-defined labeling functions. 
Current methods like WeSAL~\cite{nashaat2020wesal} ~\cite{Asterisk} ~\cite{cohen2019interactive} are based on a static set of labeling functions, 
which limits their capability make full use of the query budget for further exploration.
In addition, most of them face difficulties to deal with the extreme class imbalance found in ontology matching.
Recently ReGAL~\cite{kartchnerregal} and 
IWS~\cite{IWS} provide the possibility to 
propose additional rule-based labeling functions. 
They require users to provide feedback directly on the quality of the new functions, which becomes highly nontrivial for advanced labeling functions based on complex machine learning models, including the ones we apply in our work.
In contrast to such methods, DualLoop can augment the existing labeling function committee with additional functions that are tuned automatically for effective exploration of the search space, and, within the committee, the set of already annotated data samples is used to estimate the function quality.
\section{Conclusion and Future Work}

In this work, we introduced a novel active learning-based method named \emph{DualLoop} for identifying matched classes between ontologies. To address the challenge of highly imbalanced data in such matching tasks, DualLoop introduces two cooperative active learners: a short-term learner in a fast loop for querying user feedback on likely matching candidates predicted by an ensemble of heuristics (exploitation), and long-term learners in a slow loop to explore the space of potential matches by automatically creating and tuning new heuristics (exploration).

Our evaluation results across three datasets encompassing multiple matching tasks demonstrate that DualLoop can surmount the limitations of existing methods, consistently achieving higher F1 scores and recall compared to all baselines. Furthermore, it significantly reduces the expected query cost required to identify 90\% of all matches by more than 50\%.

We showcased the utilization of DualLoop in a commercial toolkit called TrioNet and presented its performance within an industrial use case, illustrating the applicability and benefits of our proposed method over existing tools. In the future, we plan to enhance its performance by integrating new types of tunable heuristics, such as functions based on graph embedding features or transitivity-based reasoning. 
\bibliographystyle{ACM-Reference-Format}
\bibliography{ref}

\end{document}